\documentclass[conference]{IEEEtran}
\IEEEoverridecommandlockouts
\usepackage{cite}
\usepackage{amsmath,amssymb,amsfonts}
\usepackage{algorithmic}
\usepackage{graphicx}
\usepackage{textcomp}
\usepackage{xcolor}
\usepackage{booktabs}
\usepackage{multirow}
\usepackage[T1]{fontenc}
\usepackage[utf8]{inputenc}
\usepackage{authblk}
\usepackage{marvosym}
\def\BibTeX{{\rm B\kern-.05em{\sc i\kern-.025em b}\kern-.08em
    T\kern-.1667em\lower.7ex\hbox{E}\kern-.125emX}}
\begin{document}

\title{A Pilot Study of Relating MYCN-Gene Amplification with Neuroblastoma-Patient CT Scans

\thanks{* Correspondence to {\tt xex@hust.edu.cn} (Xiang Xiang)}
}


\author[1]{Zihan Zhang}
\author[*1]{Xiang Xiang }
\author[2]{Xuehua Peng}
\author[2]{Jianbo Shao}
\affil[1]{MoE Key Lab of Image Info Processing $\&$ Intelligent Control, School of Artificial Intelligence $\&$ Automation,}
\affil[2]{Wuhan Children's Hospital affiliated with Tongji Medical College,}
\affil[1,2]{Huazhong University of Science and Technology, Wuhan 430074, China}
\renewcommand\Authands{ and }

\maketitle

\begin{abstract}
Neuroblastoma is one of the most common cancers in infants, and the initial diagnosis of this disease is difficult. At present, the MYCN gene amplification (MNA) status is detected by invasive pathological examination of tumor samples. This is time-consuming and may have a hidden impact on children. To handle this problem, we adopt multiple machine learning (ML) algorithms to predict the presence or absence of MYCN gene amplification. The dataset is composed of retrospective CT images of 23 neuroblastoma patients. Different from previous work, we develop the algorithm without manually-segmented primary tumors which is time-consuming and not practical. Instead, we only need the coordinate of the center point and the number of tumor slices given by a subspecialty-trained pediatric radiologist. Specifically, CNN-based method uses pre-trained convolutional neural network, and radiomics-based method extracts radiomics features. Our results show that CNN-based method outperforms the radiomics-based method.
\end{abstract}


\section{Introduction}
Neuroblastoma is one of the most common extracranial solid tumors in infant patients \cite{b1}. Despite a variety of treatment options, patients with high-risk neuroblastoma tend to have poor prognoses and low survival. MYCN amplification is detected in 20\% to 30\% of neuroblastoma patients \cite{b2}. MNA is an important part of the neuroblastoma risk stratification system. It has been proved to be an independent predictor and is related to aggressive tumor behavior and poor prognosis \cite{b3}. The MYCN gene with higher amplification multiple indicates that the neuroblastoma may be a more invasive type and its prognosis may be worse. Therefore, MNA patients of any age are "high-risk" groups \cite{b1}, and the detection of MNA is an essential part of the evaluation and treatment interventions of neuroblastoma.

MYCN gene amplification status is generally detected by invasive pathological examination of tumor samples which is time-consuming and may have a hidden impact on children. Therefore, it is significant to develop a fast and non-invasive method to predict the presence or absence of MNA. 

Radiomics \cite{b4} is a method to rapidly extract innumerable quantitative features from tomographic images. This allows the transformation of medical image data into high-dimensional feature data. Radiomics is composed of a set of first-order, second-order, and higher-order statistical features on images. Previous studies \cite{b6,b7,b8} have shown significant relationships between image features and tumor clinical features. For example, Wu et al. \cite{b6} first segment primary tumors and extract radiomics features automatically from the ROI. An ML model is then trained with selected features. 

Convolutional neural networks (CNN) are under-explored in the prediction of outcomes in neuroblastoma patients. CNN has shown incredible success in image classification tasks \cite{b9}, and it is a potential approach for processing medical images. Although CNN is primarily driven by large-scale data, transfer learning has shown its effectiveness in training models with small amounts of data \cite{b10}. The number of our CT data is limited, and we use a pre-trained CNN model to handle the challenge of lack of medical image data. 

In this work, we investigate the radiomics-based method and CNN-based method on a limited dataset. Specifically, we feed radiomics features into multiple ML models to predict the status of MNA. For CNN-based method, we use pre-trained ResNet \cite{b5} to extract deep features and predict the label of the data end-to-end. To the best of our knowledge, our method is the first study to try to simplify the annotation process. Specifically speaking, we do not need a pediatric radiologist to manually segment primary tumors which is time-consuming and not practical in clinical applications. Instead, we only need a pediatric radiologist to point out the center point of the tumor and the number of tumor slices in CT images. We crop the ROI images with fixed size and feed them into the model to predict the MNA status. This can greatly reduce the evaluation time of new CT data. Our results demonstrate comparable performance of previous segment-tumor method.

In summary, the contribution of this paper are as follows.
\begin{itemize}
	\item We propose a novel CNN-based method to predict the presence or absence of MYCN gene amplification of the CT images.
	\item We greatly simplify the annotation process which makes the prediction process fast and practical, and we have achieved comparable performance with previous works while the evaluation time is greatly reduced.
\end{itemize}

In the following, we first review related work and the clinical data preparation, then elaborate on radiomics-based and CNN-based methods, and further empirically compare them, with a tentative conclusion followed in the end.

\section{Related Work}
There is an increasing interest in the prediction of patient outcomes based on medical images \cite{b17,b18,b19,b20,b21,b22}. Wang et al. \cite{b12} propose a CNN-based method to predict the EGFR mutation status by CT scanning of lung adenocarcinoma patients. By training on a large number of CT images, the deep learning models achieve better predictive performance in both the primary cohort and the independent validation cohort. Wu et al. \cite{b6} combine clinical factors and radiomics features which are extracted from the manually delineated tumor. The combined model can predict the MNA status well. However, the annotation process makes the evaluation time-consuming. When evaluating new patient images, the method has to annotate the tumor ROI at first. Similarly \cite{b6}, Liu et al. \cite{b11} extract radiomics feature from tumor ROI and apply pre-trained VGG model to extract CNN-based feature. Angela et al. \cite{b13} sketch the ROI on the CT images of neuroblastoma, then extract the radiomics features on ROI. With the extracted feature, they develop the radiology model after feature selection to predict the MNA status.

\section{Clinical Data Preparation}
\textbf{Dataset.} \ From the medical records, a total of 23 patients with pretreatment CT scans who have neuroblastoma are selected. Each patient has three-phase CT images. Inclusion criteria are (1) age $\le$ 18 years old at the time of diagnosis, and (2) histopathologically confirmed MNA status detection. The number of presence of MNA in the enrolled patients is only two. The rest 21 patients do not have MNA. 

\textbf{Data preprocessing.} \ The unit of measurement in CT scans is the Hounsfield Unit (HU). We first transform it into the gray level. In the CT scans, a pixel spacing may be [2.5, 0.5, 0.5], which means that the distance between slices is 2.5 millimeters. And the pixel spacing of different CT scans may vary. As a result, we resample the full dataset to a certain isotropic resolution. Then we transfer the CT scans into image format.

The proportion of MNA and non-MNA in the training cohort is highly imbalanced (2:21). The imbalance harms the generalizability and fairness of the model \cite{b16}. To tackle this problem, we adopt re-sampling method to augment the MNA CT image data. Specifically, we apply rotation, flipping, noise injection, and gamma calibration transformation techniques to CT images. For non-MNA images, we randomly select transformation techniques to augment the images, and for MNA images, we apply all transformation techniques to balance the dataset.

With the annotation information, we use a fixed-size filter (128$\times$128 size) to crop the tumor out of each slice image around the center point of the tumor, and the cropped slice number is identical to the annotated tumor slice number. That ensures the extracted features correspond to the same spatial information across all images. 
\begin{figure}
\centering
\includegraphics[scale=0.2]{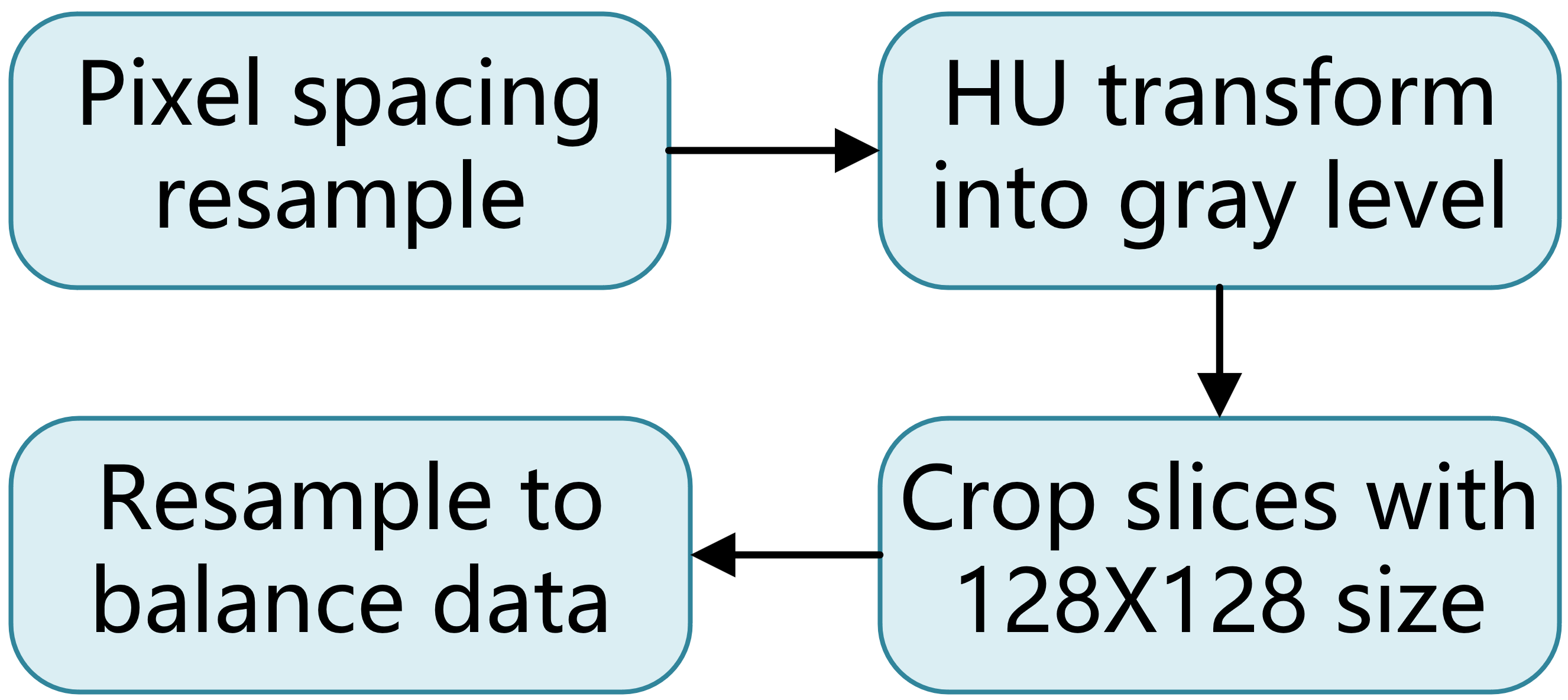}
\caption{The pre-processing of the CT data.}
\vspace{-2mm}
\label{fig:preprocessing}
\end{figure}

\section{Proposed Methods}
\subsection{Radiomics-based method}
Primary tumors are annotated from initial staging CT scans using open-source software package ITK-SNAP \cite{b14} by a subspecialty-trained pediatric radiologist. We use pyradiomics \cite{b15} to extract radiomics features, which is implemented based on consensus definitions of the Imaging Biomarkers Standardization Initiative (IBSI). We extract three kinds of radiomics features as shown in Table~\ref{table:feature}, 107 features in total. In summary, the first-order statistical features capture the intensity of the images. The shape features describe the geometric shape of the tumor. In our setting, the shape of the tumor is a cube (we do not precisely segment the tumor ROI), which may make the feature not separable because each tumor shape is similar. The gray level features represent the spatial relationship of the voxels.

After the feature extraction, we select the features which are highly  correlated to the label. In specific, we apply LASSO linear regression to select the proper features, which reduces the dimension of the data and the number of features and it attenuates over-fitting.

Finally, We adopt multiple ML methods to predict the MNA status including SVM, logistic regression, KNN, random forest and AdaBoost. The selected features and the label are the input of the model, and we train the model on CT images of 18 patients while other CT images are used for validation. We use stratified four-fold outer cross-validation to analyze the performance of our models.

\renewcommand\arraystretch{1.2}
\begin{table}
\begin{center}
\caption{Extracted radiomics features}
\label{table:feature}
\setlength{\tabcolsep}{4mm}{
\begin{tabular}{ccc}
    \toprule
     First-order statistics  & Shape-based  &  Gray Level \\
    \noalign{\smallskip}
    \hline
    \noalign{\smallskip}
    Range    & \multirow{2}{*}{2D Shape features} & GLCM\\
    Maximum  &     &   GLDM  \\
    Minimum  & \multirow{3}{*}{3D Shape features} & GLRLM\\
    Mean     &     &   GLSZM  \\
    Variance &     &   NGTDM  \\
    \hline
    Total 18 & Total 14 & Total 75 \\
    \bottomrule
\end{tabular}}
\end{center}
\end{table}

\begin{figure}
\centering
\includegraphics[scale=0.25]{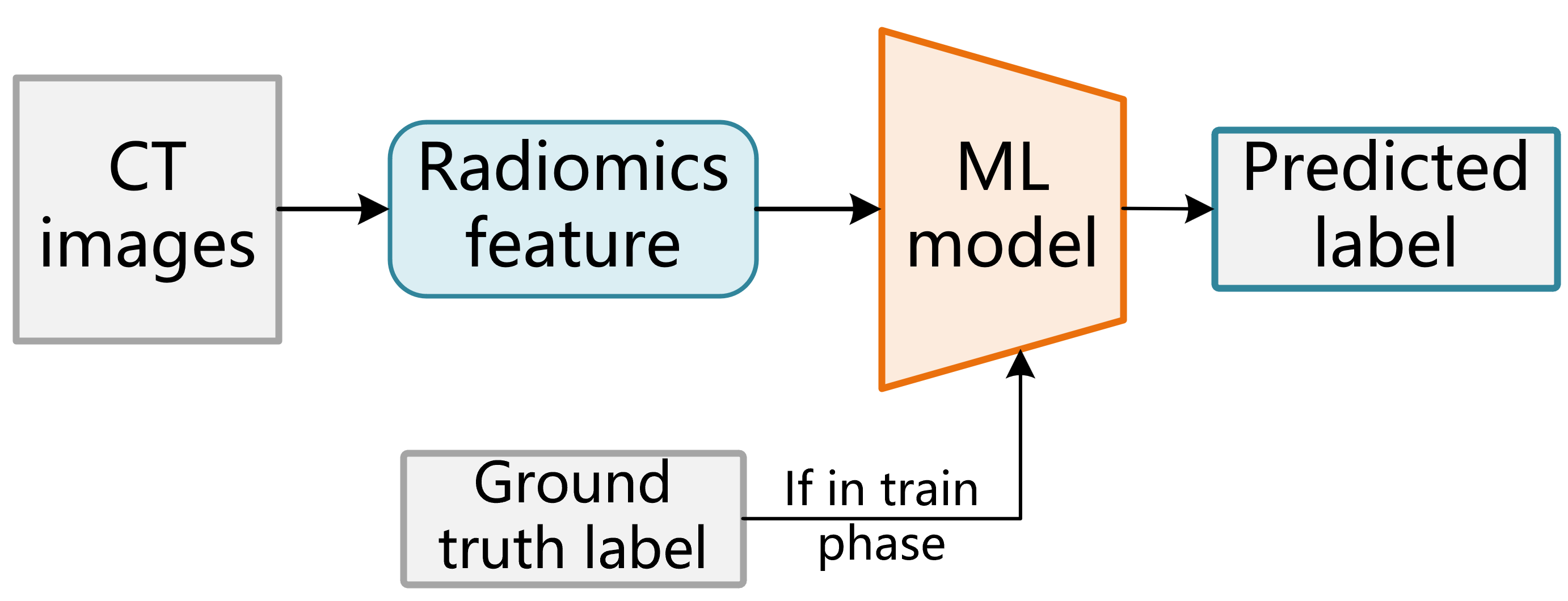}
\caption{Illustration of radiomics-based method process.}
\vspace{-2mm}
\label{fig:radiomics}
\end{figure}

\subsection{CNN-based method}
ResNet is widely used in computer vision. We adopt ResNet34 which is pre-trained on ImageNet to predict patient MNA status end-to-end from the CT images. We retrain the final layer of ResNet34 to predict the MNA status. To fit in the input size of the model, we crop the images into 128$\times$128 size based on the center point of the tumor to ensure the tumor is at the center of the cropped images. The CT images are gray images while the model requires RGB images which are three-channel. We study three approaches to transform gray images into three-channel images. In specific, the first approach inputs the gray images into the model. The second approach transforms identical gray images into three-channel images. The third approach transforms gray images of adjacent slices into three-channel images.

\begin{table}
\begin{center}
\caption{Accuracy of machine learning models for MNA status prediction. The number in brackets is the variance  and the outer number is the mean of four-fold cross-validation accuracy. The best ROC-AUC value is 0.84(95\% CI: (0.81, 0.86))}
\label{table:Ra_result}
\begin{tabular}{c|c}
    \toprule
    ML methods & Accuracy\\
    \noalign{\smallskip}
    \hline
    \noalign{\smallskip}
    SVM  &   0.73\,($\pm$0.11)\\
    Logistic regression & \textbf{0.74\,($\pm$0.11)}\\
    KNN &  0.72\,($\pm$0.09)\\
    Random forest & 0.71\,($\pm$0.09)\\
    AdaBoost &  0.70\,($\pm$0.12)\\
    \bottomrule
\end{tabular}
\end{center}
\end{table}

\section{Experiments}
We compare the performance of radiomics-based ML methods and CNN-based methods on our dataset. 

\subsection{Experimental Results}
For all experiments, we split the dataset into training set and validation set and do four-fold cross-validation. The training set contains 18 patients CT images while the validation set contains 5 patients CT images. The accuracy is reported on the total validation set images.

\textbf{Radiomics-based methods.}\ Among the ML techniques we experiment with, logistic regression model over radiomics features outperforms other models predicting MNA status as shown in Table~\ref{table:Ra_result}. The mean accuracy of logistic regression model is 0.74, 0.01 higher than SVM model. We further present the ROC-AUC value of the best logistic regression model, which is 0.84\,(95\% CI: (0.81, 0.86)). 

\textbf{CNN-based methods.}\ As shown in Table~\ref{table:All_result}, the CNN-based methods outperform the best result of radiomics-based method. The second CNN-based method whose input is synthesized by three identical gray images achieves the best performance 0.79 accuracy and ROC-AUC value 0.87. The ACC result is 0.05 higher than the radiomics-based methods.

\begin{table}
\begin{center}
\caption{Mean accuracy of radiomics-based methods and CNN-based methods for MNA status prediction. FS: Feature Selection. ACC: Accuracy}
\label{table:All_result}
\begin{tabular}{ccc}
    \toprule
    Methods-based  &  ACC & AUC\\
    \noalign{\smallskip}
    \hline
    \noalign{\smallskip}
    Radiomics-based & 0.72 & 0.81 \\
    FS + Radiomics-based  & 0.74 & 0.84\\
    $1_{st}$ CNN-based & 0.73 & 0.85 \\
    $2_{nd}$ CNN-based & \textbf{0.79} & \textbf{0.87} \\
    $3_{rd}$ CNN-based & 0.73 & 0.83 \\
    \bottomrule
\end{tabular}
\end{center}
\end{table}

\subsection{Discussion}
In this study, we investigate multiple methods to predict the MNA status based on the CT scans of neuroblastoma patients. A total of 23 patients are enrolled with MNA detection report. To the best of our knowledge, there is no such study in the analysis of CT images in neuroblastoma. 

\textbf{Radiomics-based methods.}\ In Table~\ref{table:Ra_result}, we notice that there is no significant difference in the performance of different ML models. The mean accuracy of the logistic regression model is just 0.04 higher than the AdaBoost model. The results reported in \cite{b6} are higher than ours because we report the results on the total validation images rather than the patients. Specifically speaking, we test our model on each tumor slice image and report the accuracy rather than test the model on each patient. If the output of our model is the same as the validation patients number, the mean accuracy of our radiomics-based methods is 0.882 which is 0.06 higher than the 0.826 reported in \cite{b6}. In addition, we observe a performance promotion of feature selection as shown in Table~\ref{table:All_result}. When using radiomics-based methods without feature selection, the mean accuracy is 0.72 while with feature selection, the mean accuracy is 0.74. That demonstrates the effectiveness of feature selection which helps the model to focus on the important features. 

\textbf{CNN-based methods.} \ Compared to radiomics-based methods, the CNN-based methods achieve higher performance both on the accuracy and AUC. The mean accuracy of CNN-based methods is 0.79 which is 0.05 higher than the best radiomics-based methods.

We notice that, in \cite{b11}, the results are partly opposite to the results drawn from our experiments. This
is likely because of the following reason. \cite{b11} uses precisely annotated tumor ROI to extract 3D radiomics features to predict patient outcomes. The 3D features contain more information including the size and shape of the tumor, which helps much to the prediction process. Instead, we do not need the time-consuming segmentation of primal tumors, and CNN focuses on the information of 2D images and performs better on fixed-size images. 

As shown in Table~\ref{table:All_result}, we study three approaches to transform gray images into three-channel images. The second approach performs best, and the mean accuracy is 0.06 higher than other approaches. We use the third approach that synthesizes gray images of adjacent slices to three-channel images to capture inter-slices information. However, the performance of this method is worse than the second one. This may be because the original image contains enough information to predict the MNA status.

\section{Conclusion}
Our study provides insight into that the CNN model has the capability to  perform well in the prediction of MNA status of neuroblastoma patient CT scans. In our experiments, the CNN model outperforms multiple radiomics-based ML methods. Different from previous works, we study a much less time-consuming annotation approach which greatly reduces the validation time without manually segmenting primary tumors. We also investigate different approaches to synthesize three-channel images by the original gray images and we find that duplicating the slice image into three-channel images performs better.

Due to the computational limitations, we could not perform a study to investigate more CNN models including 3D CNN which may capture inter-slices information better. Also, the radiomics-based methods in our setting is not fully explored. We hope these will inspire future work.

\bibliographystyle{splncs04}


\end{document}